\documentclass[a4paper]{jpconf}
\usepackage{graphicx}
\begin{document}
\title{Theories for the Fermi Scale}

\author{Gian Francesco Giudice}

\address{Theory Division, CERN, Geneva, Switzerland}

\ead{gian.giudice@cern.ch}

\begin{abstract}
I give a short review of our present understanding of new theories of the electroweak scale, with emphasis on recent progress.
\end{abstract}

The understanding of particle dynamics at the Fermi scale and its relation to the mechanism of electroweak (EW) symmetry breaking has been a central issue in theoretical high-energy physics for decades,  and 
direct exploration of the relevant energy domain will soon start with the beginning of
the LHC operations. In this talk I will review the theoretical expectations for new physics at the Fermi scale, concentrating mainly on the most recent developments in the field. To structure the presentation, I will address ten questions which, at present, are object of theoretical speculation, but which (hopefully) will be answered by experimental searches at the LHC. 

\section{Is there a Higgs?}

Certainly the Higgs mechanism is the most economical solution for breaking EW symmetry. Moreover the fit of EW precision data is consistent with the SM, giving some indications for the presence of a light Higgs. The preferred value of the Higgs mass is $m_H=76^{+33}_{-24}$~GeV, with a 95\% CL upper limit $m_H <144$~GeV, raised to $m_H <182$~GeV once the direct lower limit $m_H >114$~GeV is included~\cite{lepewwg}.

\begin{figure}[ht]
\begin{minipage}{16pc}
\includegraphics[width=16pc]{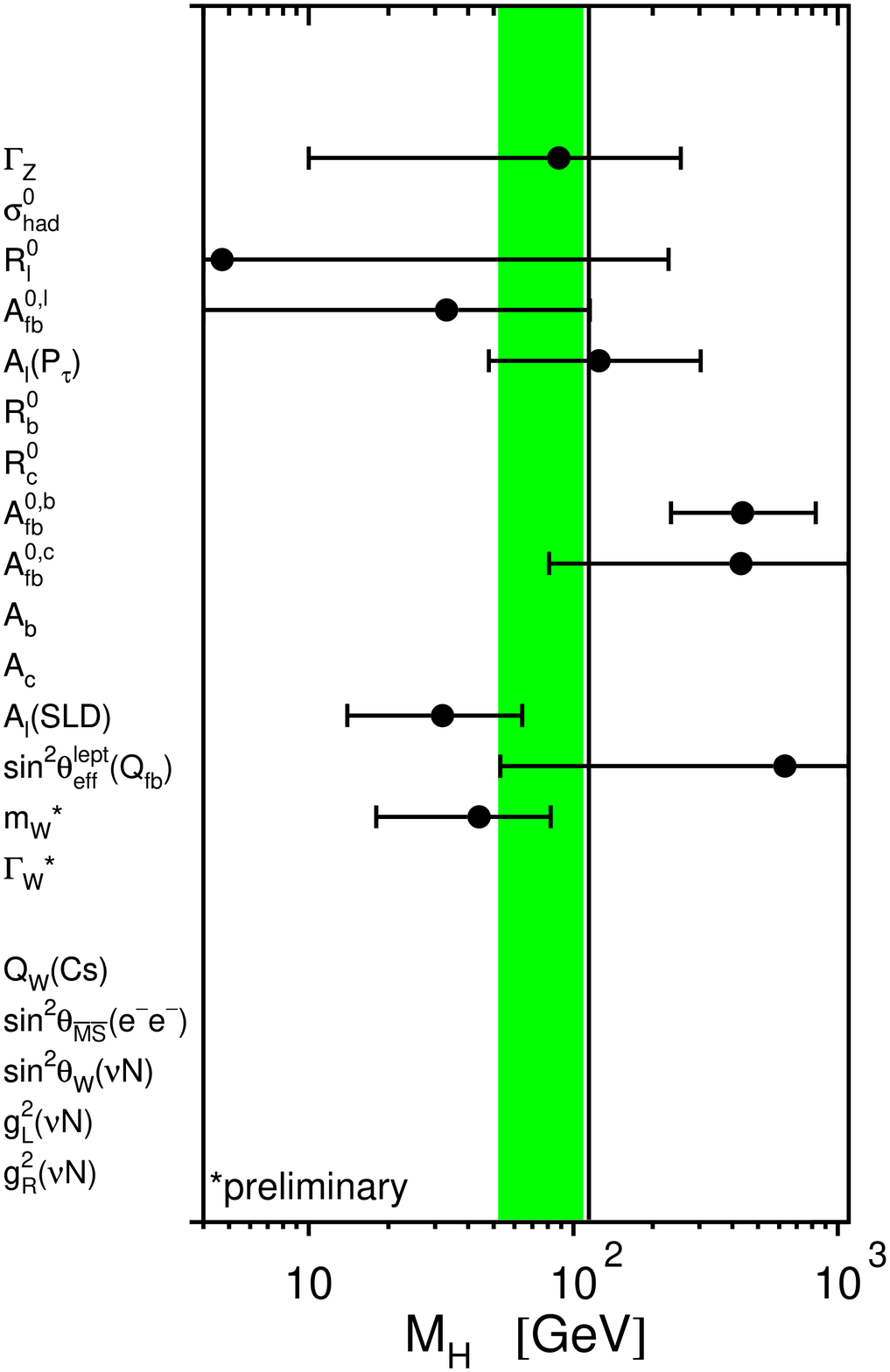}
\caption{Values of the Higgs mass extracted from different EW observables. The average is shown as a green band~\cite{lepewwg}.}
\label{fig1}
\end{minipage}\hspace{2pc}%
\begin{minipage}{20pc}
\includegraphics[width=20pc]{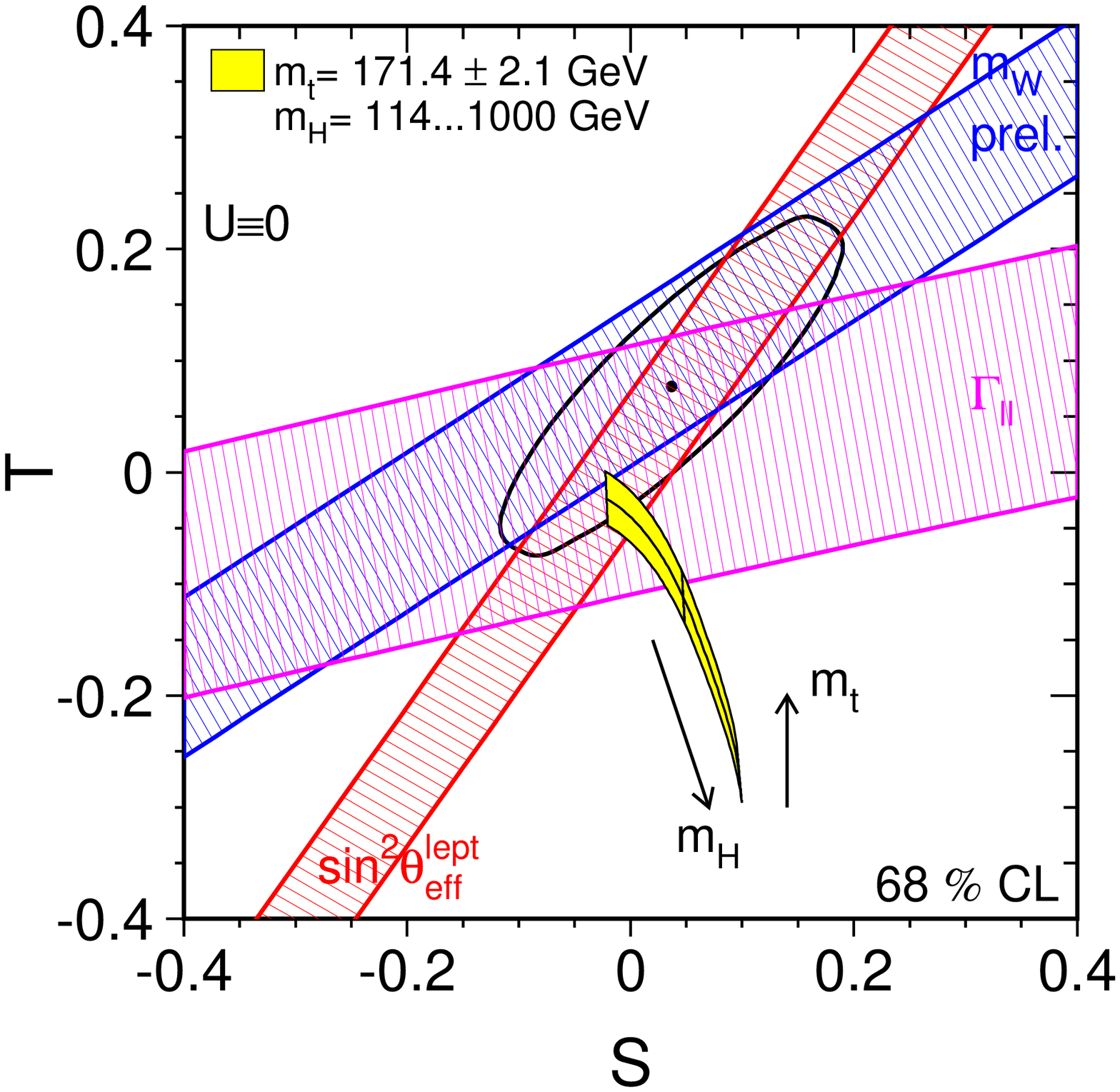}
\caption{The 1-$\sigma$ range of the EW parameters $S$ and $T$ determined by different EW observables. The ellipsis shows the 68\% probability from combined data. The yellow area gives the SM prediction with $m_t$ and $m_H$ varied as shown~\cite{lepewwg}.}
\label{fig2}
\end{minipage} 
\end{figure}

There are however some mild reasons of concern for the SM picture with a light Higgs. First of all, the decrease in the value of the top-quark mass measured at the Tevatron has worsened the SM fit, giving a $\chi^2/$dof of 18/13 corresponding to a probability of 15\%~\cite{gamb}. In particular, the value of the top mass extracted from EW data (excluding the direct Tevatron measurements) is $m_t=178.9^{+11.7}_{-8.6}$~GeV, while the latest CDF/D0 result is $m_t=170.9\pm 1.8$~GeV~\cite{tev}. Of more direct impact on the light-Higgs hypothesis is the observation that the two most precise measurements of $\sin^2\theta_W$ do not agree very well, differing by more than 3 $\sigma$. The $b\bar b$ forward-backward asymmetry ($A_{fb}^{0,l}$) measured at LEP gives a large value of $\sin^2\theta_W$, which leads to the prediction of a relatively heavy Higgs with $m_H=420^{+420}_{-190}$~GeV. On the other hand, the lepton left-right asymmetry ($A_l$) measured at SLD (in agreement with the leptonic asymmetries measured at LEP) gives a low value of $\sin^2\theta_W$, corresponding to $m_H=31^{+33}_{-19}$~GeV, in conflict with the lower limit $m_H>114$~GeV~\cite{leph} from direct LEP searches. Moreover, the world average of the $W$ mass, $m_W=80.392\pm 0.029$~GeV, is still larger than the value extracted from a SM fit, again requiring $m_H$ to be smaller than what is allowed by the LEP Higgs seaches. The situation is summarized in fig.~\ref{fig1}~\cite{lepewwg}, where the predicted values of $m_H$ from the different observables are shown.  While $A_{fb}^{0,l}$ prefers a relatively heavy Higgs, $A_l$ and $m_W$ require a very light Higgs, already excluded by LEP. Only when we average over all (partially inconsistent) data, we obtain the prediction for a relatively light Higgs and the well-known upper bound $m_H<182$~GeV. However, the fit of the observables most sensitive to $m_H$ has a probability of less than 2\%~\cite{gamb}. Although there is little doubt that the SM gives a satisfactory description of all EW data, this observation makes, in my opinion, the argument in favor of a light Higgs less compelling. 

Therefore it is still open the possibility that no light particles with the quantum numbers of the Higgs exist in nature, and that EW-symmetry breaking is triggered by a new sector with strong dynamics. Technicolor is the prototype of such theories. An interesting realization of Higgless theories is in the context of extra dimensions, where EW symmetry is broken by non-trivial boundary conditions of the gauge fields in the compactified extra coordinates~\cite{higgsless}.
The associated Kaluza-Klein modes of the gauge bosons modify the energy growth of the amplitude for longitudinal gauge-boson scattering, postponing the violation of unitarity beyond the TeV. This can be understood with a simple argument of dimensional analysis. While in a normal 4-$d$ gauge theory the violation of unitarity occurs at the scale $\Lambda_4 \simeq 4\pi m_W/g\simeq$~TeV, in a 5-$d$ theory this happens at $\Lambda_5 \simeq 24 \pi^3/g_5^2$, where the dimensionful 5-$d$ gauge coupling is related to the 4-$d$ coupling by $g_5^2=2\pi R g^2$. If the radius $R$ of the compactified 5$^{\rm th}$ dimension is associated to the inverse of the $W$ mass, then $\Lambda_5 \simeq 12 \pi^2 m_W/g^2$ could in principle be as large as about 10 TeV~\cite{barbrat}.
 
 The traditional problems with technicolor and Higgsless theories are the generation of fermion masses/mixings and EW precision data. Figure~\ref{fig2} shows the measured values of the main EW observables expressed in terms of $S$ and $T$, the coefficients of the leading form factors contributing to oblique corrections. The SM prediction and its dependence on $m_t$ and $m_H$ is also shown. In $SU(N)$ technicolor with $n$ technifermion flavors, one finds an extra contribution $\Delta S \simeq (nN+\ln \Lambda_{TC}/m_Z)/6\pi$. The logarithmic term corresponds to the infrared effect below the technicolor scale $\Lambda_{TC}$, due to the absence of a Higgs boson, while the first term corresponds to the loop contribution of technifermions in the $Z$ propagator. The total effect is to bring the theoretical prediction far away from the experimentally allowed ellipsis shown in fig.~\ref{fig2}. An additional large and negative contribution to $S$ (and a positive contribution to $T$) is required. Recently the question of generating negative contributions to $S$ has been revisited~\cite{negs}, but all calculable models in extra dimensions were shown to give positive $S$~\cite{barbrat,nonegs}. Technicolor models with technifermions transforming as larger representations than fundamentals were found to have interesting features~\cite{sannino}. They can be near the conformal window (where walking dynamics~\cite{walking} can take place and can help addressing the problem of generating a realistic quark mass spectrum) at small values of $N$ (where the positive contribution to $S$ is reduced).

\section{What is the Higgs mass?} 
  
If the Higgs boson is discovered, its mass will be one of the most important observables to be measured. Not only will it determine the Higgs quartic coupling, the last unknown parameter of the SM, but it will also give us important indirect information about possible extensions of the SM at very short distances. Experimental data already restrict the allowed values of the Higgs mass
\begin{equation}
114~{\rm GeV}<m_H<182~{\rm GeV}.
\label{wind1}
\end{equation}
The lower limit is the 95\% CL exclusion from direct LEP searches~\cite{leph}. The upper limit is less robust, since it assumes no new-physics contributions to EW data, and it is subject to the data interpretation previously discussed. 

Theoretical considerations give bounds on $m_H$ too. A lower limit is derived by requiring that the EW vacuum has a lifetime longer than the age of the Universe, and an upper bound is imposed by the absence of a Landau pole in the Higgs quartic coupling. These limits depend on the maximum energy scale $\Lambda$ at which the SM can be extrapolated, and become more stringent the larger $\Lambda$ is. For $\Lambda$ equal to an hypothetical GUT scale of $10^{16}$~GeV, one finds~\cite{limith}
\begin{equation}
125~{\rm GeV} <m_H<175~{\rm GeV}.
\label{wind2}
\end{equation}

The approximate coincidence between the experimentally and theoretically allowed windows in eqs.~(\ref{wind1})--(\ref{wind2}) can be viewed as interesting or worrisome, depending on the point of view. It is interesting that experimental and theoretical information concur to determine the Higgs mass in a relatively narrow range. It is however worrisome that, in this window of $m_H$, the SM can be extrapolated, in a fully consistent way, to extremely high-energy scales with a ``big desert" picture up to  $M_{\rm GUT}$ or even $M_{\rm Pl}$. The presence of right-handed neutrinos at some intermediate scale does not significantly change this conclusion, unless their Majorana masses are very large and we insist to extrapolate the theory at scales even beyond $M_{\rm GUT}$~\cite{spagneu}.

If the Higgs mass is found to lie outside the overlap region between  eqs.~(\ref{wind1}) and (\ref{wind2}), we can immediately infer some properties on the short-distance structure of the theory. If $m_H>182$~GeV, new dynamics at the Fermi scale is likely to modify the interpretation of the EW data and to affect the evolution of the Higgs quartic coupling. If $114~{\rm GeV}<m_H<130~{\rm GeV}$, the EW vacuum is potentially metastable, and either new physics modifies the Higgs potential at large field values or the cosmology has to satisfy appropriate constraints not to destabilize the vacuum at early times~\cite{giurio}.

The Higgs mass is also a very useful discriminator for new-physics models. Take the case of supersymmetry, where both the $Z$ and the Higgs masses are calculable in terms of soft-breaking parameters
\begin{eqnarray}
m_H^2 &=& m_Z^2\cos^22\beta +\frac{3G_Fm_t^4}{\sqrt{2}\pi^2}\ln \frac{{\tilde m}_t^2}{m_t^2}+\dots \label{higgsm} \\
m_Z^2 &=& \frac{3\sqrt{2}G_F m_t^2 {\tilde m}_t^2}{2\pi^2}\ln \frac{\Lambda^2}{{\tilde m}_t^2}+\dots .
\label{topsm}
\end{eqnarray}
As is well known, a value of $m_H$ sufficiently large to avoid the LEP bound requires heavy stops, see eq.~(\ref{higgsm}). In turn, a large stop mass ${\tilde m}_t$ gives a large contribution to $m_Z$, see eq.~(\ref{topsm}), and a certain amount of tuning among different soft-breaking parameters is needed to reproduce the physical value of $m_Z$. The situation has worsened with the decrease in the measured top-mass value, since the loop contribution to $m_H^2$ is proportional to $m_t^4$, see eq.~(\ref{higgsm}). Fixing $ {\tilde m}_t=1$~TeV and $m_t$ at its central value, we find that the case of no stop mixing is excluded, and the bound $m_H<128$~GeV is obtained for arbitrary mixing~\cite{feynhig}. However, the value of the mixing that maximizes $m_H$ is highly unstable under radiative corrections, 
and gluino effects tend to drive the mixing parameter to a fixed point, where the new contribution to the Higgs mass is very modest. For instance in gauge mediation with  $ {\tilde m}_t=1$~TeV, the Higgs mass can at best be barely above the LEP bound. However, the situation can greatly improve if the stop square masses start negative at the GUT scale~\cite{kim1}, a condition that can be satisfied, for instance, in models with gauge messengers~\cite{kim2}.
Even larger values of $m_H$ can be obtained in drastic modifications of the minimal supersymmetric models, such as in split supersymmetry~\cite{split}, or in extensions with new tree-level contributions to the Higgs quartic from extra gauge interactions or extra singlet fields coupled to the Higgs. The new couplings introduced in these extensions typically become strong at some intermediate scale and the prediction of gauge coupling unification in supersymmetry is usually lost. One can collectively describe these extensions with new Higgs quartic couplings using an effective-theory language. The leading operators arising from additional dynamics at the scale $M$ are (denoting by $S$ and $F$ the scalar and auxiliary superfield components)~\cite{opanls}
\begin{equation}
{\cal L}=\frac{1}{M}\left[ c_1\left. \left( H_uH_d\right)^2 \right|_F +c_2{\tilde m}\left. \left( H_uH_d\right)^2 \right|_S \right] +{\rm h.c.}
\end{equation}
Therefore, the Higgs interactions and mass spectrum in extensions of the minimal model with new tree-level quartic couplings is described, at leading order in $M$, by only two new coefficients ($c_{1,2}$). The LHC will tell us if minimal supersymmetry is a viable theory or if the addition of such new structures is a necessary complication. Were the experiments to reveal a Higgs boson with $m_H>130$~GeV and sub-TeV stops, we would have a clear indication for additional structure beyond minimal supersymmetry. 

\section{Is the Higgs a SM-like weak doublet?}

The choice of embedding the Higgs into a single $SU_2$ doublet is dictated only by simplicity, and the Higgs sector may contain more fields in weak doublets (like in minimal supersymmetry), weak singlets or even larger representations. There has been recent activity in studying extensions of the minimal SM Higgs structure. One motivation~\cite{port} is that the Higgs is the field most sensitive to the existence of hypothetical hidden sectors, {\it i.e.} new particles which do not carry any quantum numbers under the SM gauge group. This is because the Higgs mass term is the only super-renormalizable interaction allowed in the SM and therefore a hidden-sector dimension-2 operator $\cal O$ can have a coupling of the form 
\begin{equation}
{\cal L} = c\left| H \right|^2 {\cal O},
\label{effint}
\end{equation}
without any suppression of inverse powers of large masses. Once the Higgs gets an EW-breaking vev, it can mix with singlet states present in the hidden sector. 

A second motivation for studying extended Higgs sectors is to evade the upper bound on $m_H$ coming from the usual interpretation of EW data. As shown in fig.~\ref{fig2}, one can compensate the effect of a heavy Higgs with extra contributions giving small $\Delta S$ and $\Delta T \simeq 0.3$, which bring the theoretical prediction back inside the experimentally allowed ellipsis. There are several interesting options to achieve this goal. A simple possibility is to enlarge the Higgs sector to contain two Higgs doublets $H_{1,2}$ with a discrete parity, under which $H_2$ is odd, choosing the potential such that $\langle H_2\rangle=0$ (the ``Inert Doublet Model"~\cite{inert}). It is possible to fix the coupling constants in the scalar potential such that $\Delta T$ is of the required size, even when the Higgs is as heavy as 800~GeV. Moreover, the lightest parity-odd Higgs can be a suitable dark-matter  candidate. The same result can also be obtained~\cite{barb2} in a supersymmetric model with a new singlet superfield with superpotential $W=\lambda N H_1 H_2 +k N^3$. The desired $\Delta T$ contribution now comes from a combination of Higgs-higgsino effects. A very heavy Higgs is then allowed, although the price is that the coupling $\lambda$ blows up very early, requiring a theory cutoff typically at about 10~TeV.  New EW fermions can also reproduce $\Delta T \simeq 0.3$, without the need to extend the Higgs sector~\cite{hallew}.

A third motivation is to evade the lower bound on $m_H$ from direct LEP searches. This is especially interesting in view of the 2.3$\sigma$ excess with respect to background observed by the LEP experiments  in Higgs searches corresponding to $m_H=98$~GeV. For comparison, notice that the controversial Higgs signal at 115~GeV has an excess with respect to background of less than 2$\sigma$. In order to interpret the excess at 98~GeV as Higgs production, one needs
\begin{equation}
\left( \frac{g_{hZZ}}{g_{hZZ}^{\rm (SM)}}\right)^2 BR\left( h\to b\bar b\right) \simeq 0.2,
\end{equation}
where $g_{hZZ}/g_{hZZ}^{\rm (SM)}$ is the Higgs coupling to the $Z$ in units of the SM coupling. Therefore, we can either reduce the Higgs branching ratio $BR ( h\to b\bar b )$ by introducing new decay modes (which requires new light particles) or we can reduce the coupling to the $Z$ (which requires changing the nature of the Higgs by mixing it with a new light state). In the latter case, the state which is ``mostly Higgs" will still respect the LEP bound, while the 98-GeV mass particle will have only a 20\% Higgs component. The LEP collaborations have searched for a variety of Higgs decay modes, including Higgs decaying dominantly into $\gamma \gamma$ (with a limit $m_H>117$~GeV), into two jets ($m_H>113$~GeV), into an invisible final state ($m_H>114$~GeV), and into $b\bar b b\bar b$ ($m_H>110$~GeV). However, a totally inclusive search of $h\to$~anything was performed only by OPAL~\cite{opal}, deriving a limit of $m_H>82$~GeV. The same experiment has also obtained the limit $m_H>87$~GeV for a Higgs dominantly decaying as $h\to 4\tau$. This leaves open the possibility that a Higgs with unusual decay modes has escaped detection at LEP, in spite of having a mass in the range between 82 and 114~GeV.

Both possibilities of modifying either $g_{hZZ}$ or $BR ( h\to b\bar b )$ can actually be achieved in supersymmetry. One can find supersymmetric models where the lightest Higgs has a 98~GeV mass and reduced couplings to the $Z$, while the Higgs with nearly-SM couplings has a mass of 115~GeV~\cite{modfh}. Alternatively, there are supersymmetric models with an extra singlet with dominant Higgs decay $h\to aa$, where $a$ is a pseudoscalar with mass below the bottom threshold decaying $a \to \tau^+\tau^-$ and $c\bar c$~\cite{dergun,chang}. Another option is to assume that the Higgs primarily decays into a neutralino pair, each decaying into three quarks through an $R$-parity violating interaction~\cite{kapl}. In this case, the Higgs production at the LHC can lead to displaced vertices, a topology which could be efficiently investigated also by LHCb.

\section{Is the Higgs elementary or composite?}
If the Higgs boson is discovered, we will need to determine the nature of the force responsible for generating the EW breaking medium permeating the vacuum. In the case of the SM (with relatively light Higgs) and of supersymmetry, this force is weak and the Higgs is an elementary particle. Indeed, the case of supersymmetry is particularly emblematic. Here the Higgs shares the same nature of quarks and leptons, since they are all described by chiral superfields, corresponding to elementary particles. Moreover, there is no new force, beside gauge and Yukawa interactions, to be added to break EW symmetry.  

On the other hand, there are various interesting models where the Higgs is a composite particle, the light remnant of a strong force. The old proposal of the Higgs as a pseudoGolstone boson~\cite{georkap} is a prototype of this class of theories, but in the last few years new ideas have been pursued: the Little Higgs~\cite{littleh}, where the Higgs mass is protected by multiple approximate symmetries and it can be generated only after ``collective breaking" at two or higher loops; the gauge-Higgs unification~\cite{gaugeh,gaugeh2}, where a gauge symmetry protects the Higgs mass, since the Higgs is part of a gauge field in higher-dimensional theories; the holographic Higgs~\cite{hologh}, a realization of the previous idea with a 4-$d$ interpretation through duality with theories with a warped extra dimension. 

The distinctive feature of these theories at the LHC will be the production of new states (either the Little Higgs partners of the $W$, $Z$, $t$, or the Kaluza-Klein excitations, or the ``hadronic" resonances of the new strong force) characteristic of the underlying model. Nevertheless, it is useful to identify model-independent features of Higgs compositeness at the LHC, especially in cases in which it is not experimentally straightforward to relate the new discovery to the mechanism of EW breaking, or even in the more pessimistic case in which the new states lie beyond the LHC kinematical reach.
There has been recent progress in determining the model-independent features of a composite Higgs using an effective-theory approach~\cite{eff1,eff2,eff3}. The Goldstone-like Higgs is described by a $\sigma$-model deformed by the SM gauge and Yukawa interactions. This defines a specific pattern of modified Higgs couplings with respect to their SM value. The leading effects are described by two parameters (one for a universal modification of all Higgs couplings, and the other one for a universal modification of Higgs couplings to fermions) characterized by the ratio $v^2/f^2$, where $v$ is the Higgs vev and $f$ is the $\sigma$-model scale. Experiments at the LHC will be able to measure Higgs production rates times branching ratio with a precision of 20--40\%~\cite{lhcp}, which is often limited by statistics and therefore it can be improved up to about 10\% at a luminosity upgrading like the SLHC. The ILC could improve these measurements at the percent level~\cite{ilcp}, giving a conclusive test of Higgs compositeness up to a scale of about $4\pi f=30$~TeV.  

A genuine test of Higgs compositeness can be done at the LHC by analyzing longitudinal gauge-boson scattering. Indeed, because of the modified Higgs couplings, longitudinal gauge-boson ($V_L$) scattering amplitudes violate unitarity at high energy, in spite of the presence of a light Higgs~\cite{eff2}. The $E^2$-growing amplitude is a factor $v^2/f^2$  smaller than in the Higgsless case ($\not \! H$) and we find, in the high-energy limit,
\begin{equation}
\sigma\left( pp\to V_L V^\prime_L X\right) =\frac{v^4}{f^4}\sigma\left( pp\to V_L V^\prime_L X\right)_{\not \! H} .
\end{equation}
Therefore, even if a light Higgs is discovered, the $V_LV_L$ scattering is an important process to study, which can give us useful information on the nature of the Higgs boson. Moreover, since in these theories the Higgs can be viewed as an approximate fourth Goldstone boson, its properties are related to those of the exact (eaten) Goldstone bosons. Strong gauge-boson scattering will be accompanied by strong Higgs pair production~\cite{eff2}.

\section{Is the hierarchy $M_W/M_P$ explained by a symmetry or a dynamical principle?}

Most of the constructions of new-physics theories at the Fermi scale have focussed on the hierarchy problem as their primary motivation. Indeed, there are many examples in physics where unexpected and precise parameter cancellations were actually the signal of the existence of new particles (although in some cases the new particles were actually discovered before the problem was realized). In classical electrodynamics and in quantum mechanics the electron self-energy has a power divergence, which leads to a naturalness problem, cured only by the introduction of the positron in field theory. The electromagnetic contribution to the $\pi^+$-$\pi^0$ mass difference needs to be cut-off at the scale of the $\rho$ meson. The existence of the charm was predicted by finding a justification for the cancellation in the $K_L$--$K_S$ mass difference (an idea rewarded with this year's Dirac medal). The top-quark discovery at the EW scale was not unexpected, as this particle is needed to cancel the SM gauge anomaly. Therefore there are good precedents and credible theoretical reasons to expect that also the miraculously precise cancellation in the quantum corrections to $M_W/M_P$ will actually have a symmetry or dynamical explanation at the Fermi scale.

In the last few years an opposing view has started to emerge, fueled by experimental and theoretical considerations. Data suggesting the existence of a cosmological constant of typical size $10^{-3}$~eV and the lack of any good symmetry explanation for the smallness of this value appear to show that power-divergent quantum corrections may not necessarily be regulated by a dynamical principle. After LEP1 and LEP2, all known beyond-SM theories suffer from a mild fine-tuning problem which, although much less severe than the original hierarchy problem, makes them not fully natural. Finally, the emergence of the landscape~\cite{lands} in string theory has provided a natural setting for statistical, rather than dynamical, explanations of parameter determination. For all these reasons, physicists have tried  to pursue new frameworks which challenge our intuition based on effective theories, which is at the origin of the hierarchy problem. The most well-defined of these schemes is a selection criteria of multi-vacua theories determined by the anthropic principle, which constrains the Higgs vev within a narrow range~\cite{anthv}.

In case any of these schemes is operative to explain the hierarchy problem, one of the most pressing questions is whether we will be able to determine it with experimental observations. Although it will be generally difficult, there are cases in which the LHC can give us indications for ``unnaturalness" at work at the Fermi scale. One example is Split Supersymmetry~\cite{split}, where squarks and sleptons are made heavy, maintaining the predictions of gauge-coupling unification and dark matter, but discarding a too light Higgs, fast proton decay and the flavour problem. Another example is the explanation of the little hierarchy in supersymmetry based on a distribution of vacua growing with the supersymmetry-breaking mass, which predicts an average value of $M_Z^2/M_{\rm susy}^2$ equal to a loop factor~\cite{stat}. Another interesting observation~\cite{hallwat} is that, if the distribution of vacua grows at small $\lambda$ (the Higgs quartic coupling), the request that the EW breaking vev is within the narrow anthropic window determines the Higgs mass to be near its metastability limit $m_H=115\pm 6$~GeV.  If also the top Yukawa coupling can take any value in the landscape, the same argument determines $m_t$ in the range between 172 and 177~GeV.

\begin{center}
\begin{figure}[ht]
\hspace{4pc}
\includegraphics[width=26pc]{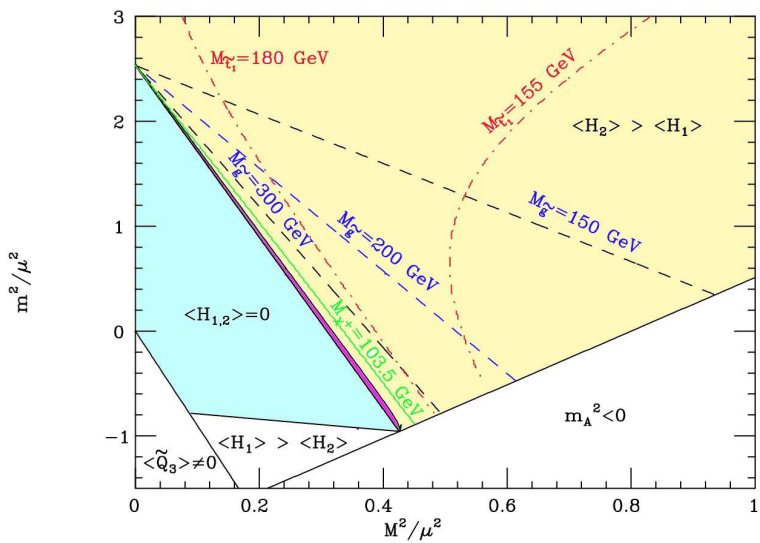}
\caption{The phase diagram of a minimal supersymmetric model with universal scalar mass $m$, unified gaugino mass $M$ and Higgsino mass $\mu$ at the GUT scale~\cite{stat}.}
\label{fig3}
\end{figure}
\end{center}

\section{Is supersymmetry effective at the Fermi scale?}

Among known solutions to the hierarchy problem with a valid extrapolation up to energies as high as the GUT mass, supersymmetry appears to be the most satisfactory. Its main problem is that neither the Higgs nor any supersymmetric particles have been observed at LEP. There are different ways of quantifying how serious this problem is. One way is illustrated in fig.~\ref{fig3}, which shows the phase diagram of a typical supersymmetric model.
In a large fraction of parameter space (the yellow area) we find a phase with symmetry breaking $SU_2\times U_1 \to U_1$ , showing that the radiative EW breaking phenomenon is a rather typical feature of low-energy supersymmetry. However, in most of this region, the supersymmetric particles have masses not far from $M_Z$ and they have been excluded by experimental searches. Only a thin sliver of parameter space survives (the purple area), a measure of the amount of tuning that supersymmetric theories have to suffer to pass the experimental tests. The surviving region has the characteristic of lying very close to the critical line that separates the phases with broken and unbroken EW symmetry. Either we are on the wrong track about supersymmetry or, if it is eventually discovered at the LHC, we will have to understand why it lies in a ``near-critical" condition with respect to EW breaking.

In the last decades we have learnt that supersymmetry can be realized at the Fermi scale in a variety of ways, each one with very different experimental signatures at the LHC. Each scheme has its advantages and drawbacks and none has emerged as the theoretically preferred. Supergravity offers an intriguing connection between EW physics and gravity, but it often leads to a large arbitrariness in the choice of the soft-breaking terms which generally amounts to excessive flavour and CP violation. While the neutralino is the most popular candidate to be the lightest supersymmetric particle (LSP), there has been a revival of the possibility that the gravitino ($\tilde G$) is actually the LSP, being also a viable dark-matter candidate when the stau is the NLSP~\cite{gravdm}. The stau would be revealed at the LHC as a long-lived or stable charged particle with distinctive time-of-flight and energy-loss signatures~\cite{stopst}. Metastable staus could be captured by ``stopper" detectors placed inside the ATLAS or CMS caverns, where one could measure the position and time of the stopped $\tilde{\tau}$, and later the time and energy of the emitted $\tau$ from the decay ${\tilde \tau} \to \tau {\tilde G}$. In this way, we could have a unique opportunity to reconstruct the original supersymmetry-breaking scale and the gravitational coupling determining the decay process. With large statistics, the gravitino spin could also be measured from ${\tilde \tau} \to \tau \gamma {\tilde G}$ distributions~\cite{buch}.

Gauge mediation is a very attractive scenario with computable soft terms free from large flavor violations. On the other hand, the origin of the $\mu$ term is more problematic (for some recent attempts to address this issue, see refs.~\cite{slav,muconf}) and the LEP bound on the Higgs mass is forcing an especially high degree of fine tuning on the theory. Indeed, the soft terms are highly constrained by the theory and this prevents the possibility of achieving large mixings in the stop sector, which can alleviate the tuning giving an extra contribution to the Higgs mass.  From the point of view of gauge-mediation model building, there has been a burst of activity triggered by the realization~\cite{iss} that, in a broad class of supersymmetric QCD with massive flavors, the origin of the potential provides a metastable susy-breaking vacuum, although the global minimum preserves supersymmetry. This, as previously suggested in a more restricted class of theories~\cite{dim}, can alleviate the problem of the restoration of supersymmetry when the hidden sector is coupled to the messengers. Many new models based on metastable susy-breaking vacua have been proposed (and the list of references is too long to be included here).

Anomaly mediation is an unavoidable effect always present in the visible sector, once supersymmetry is broken. Generally it gives only subleading effects, when other mediation mechanisms are present. However, there are cases in which anomaly mediation gives the leading contribution to soft terms: in presence of extra-dimensional separation between visible and hidden sectors~\cite{anmed1}, when there are no gauge singlets in the hidden sector~\cite{anmed2}, or in the case of conformal sequestering~\cite{confseq}. The soft terms are computable and UV insensitive, avoiding difficulties with excessive flavor violations. However, another source of supersymmetry breaking in the soft terms is needed to cure the negative slepton square masses. An interesting option is mirage mediation~\cite{mirage}, a combination of anomaly and modulus mediation, where the $F$ vev of the modulus is parametrically suppressed with respect to $m_{3/2}$, and numerically of the same order of the anomaly contribution. Although there are problems to justify the theoretical structure~\cite{jtha} (see however ref.~\cite{mir2}), mirage mediation leads to soft terms with appealing phenomenological properties. The name ``mirage" comes from the property of the soft terms to unify at an intermediate scale, which nevertheless does not correspond to any physical threshold. This mirage scale can be chosen to be $M_{mir}=M_{GUT} m_{3/2}/M_{Pl}$, which is expected to be close to a TeV, since the size of the soft terms is a one-loop factor smaller than $m_{3/2}$. At the mirage scale, the gaugino masses ($M$), the scalar masses ($\tilde m$) and the trilinear terms ($A$) satisfy a unification relation $M=A={\tilde m}\sqrt{2}$. This gives a compressed spectrum of squark and slepton masses, a small logarithmic evolution from the unification ({\it i.e.} mirage) scale, and large $A$ terms in the stop sector. These three properties are exactly what is needed to make the Higgs sufficiently heavy with less apparent tuning~\cite{mir2,spec}.

If supersymmetry is discovered, a difficult task for the experiments will be to disentangle the various theoretical possibilities and identify the pattern of soft terms. Truly model-independent analyses involve a very large number of free parameters. New strategies to go from experimental signals to theoretical interpretation have started to be investigated~\cite{newan}. Not only can this problem be experimentally challenging, but it can also be theoretically intricate. If supersymmetric particles are detected and the corresponding parameters measured, it will be interesting to extrapolate these parameters to high-energy scales, in order to compare them with theoretical predictions made at the GUT or the messenger scales. In this extrapolation, the evolution of the soft terms can be affected by hidden-sector fields , if this sector is strongly interacting~\cite{schmal}. Therefore, scalar mass unification can occur in simple theories but, in some cases, it could be impossible to derive it from collider-energy measurements because it is disguised by strong hidden-sector dynamics. On the other hand, if simple high-energy patterns of soft terms emerge at high energies, this could be interpreted as evidence that the hidden sector is weakly interacting, as in the familiar case of moduli fields. 

\section{Are there extra dimensions? Are there new strong forces?}          

Technicolor and extra dimensions are appealing scenarios to address the hierarchy problem which, at first sight, appear to be quite unrelated to each other. However, the AdS/CFT correspondence~\cite{ads} has suggested a deep relation between theories with 5-dimensional warped gravity with SM fermions and gauge bosons in the bulk and the Higgs on a brane, on one side, and walking technicolor theories with slowly-running couplings in 4 dimensions, on the other side. As almost a century ago we learnt that particles and waves are only different aspects of the same physical reality, in the same way the familiar concepts of space dimension and force may not be distinct, but just dual descriptions of the same phenomenon. The 5-dimensional picture of a TeV brane and a Planck brane separated along the 5$^{\rm th}$-dimensional coordinate can be replaced by the familiar renormalization-group flow between an infrared (TeV) and ultraviolet (Planck) scale in 4 dimensions. In this sense, there is a correspondence  also between position and energy, which is typical of a gravitational field, even in the Newtonian theory. In the case of the Randall-Sundrum setup~\cite{rs}, the energy associated to a particle changes exponentially as we move in warped space, and the familiar gravitational red-shift is so powerful that it can explain the Fermi--Planck hierarchy.
   
Indeed, I believe that a Randall-Sundrum variation with gauge bosons and fermions in the bulk is emerging as one of the most interesting models with extra dimensions (or with new forces?). It can address the hierarchy problem, be consistent with gauge-coupling unification, suppress unwanted flavour violations, and potentially explain large hierarchies in the structure of the fermion masses. Recently, the collider signatures of the theory have been revisited, taking into account the effect of appropriately locating fermions in the extra dimension to obtain a correct pattern of quark and lepton masses. The discovery channel at the LHC is resonant production of the first excited Kaluza-Klein gluon decaying into a $t\bar t$ pair, which can be identified up to a mass of about 5~TeV~\cite{gluon}. The final test of the gravitational nature of the underlying theory will come from resonant production of the Kaluza-Klein graviton, identifying its spin-2 structure. Observing  $t\bar t$ and $ZZ$ final states, LHC can discover the Kaluza-Klein graviton, up to masses of about 2~TeV~\cite{gra}.

\section{Will Dark Matter be discovered at the LHC?}

It is impossible to overestimate the importance of discovering  Dark Matter (DM) at the LHC. Such a discovery will imply a revision of the SM, it will strenghten the connection between particle physics, cosmology and astrophysics, and it will enormously enlarge our understanding of the present and past universe. Confidence that the LHC could indeed produce DM follows from the celebrated relation between the present energy density (in units of the critical density) of a thermal-relic particle ($\Omega_{DM}$) and the thermal average of its annihilation cross section times velocity ($\sigma$)
\begin{equation}
\Omega_{DM} =0.22~\frac{\rm pb}{\sigma}.
\label{om}
\end{equation}
The value $ \Omega_{DM} =0.22$, obtained by cosmological observations~\cite{wmap}, is reproduced for a pb cross section which, by simple dimensional analysis, is typical of a particle with Fermi-scale mass. Note that the number 0.22 in eq.~(\ref{om}) comes from a combination of cosmological quantities (Hubble rate, CMB temperature, number of thermal degrees of freedom, Newton's constant) but it is not parametrically related to the weak scale. The connection between thermal-relic DM and the Fermi scale appears as a numerical coincidence, but a coincidence that could be pregnant with consequences for EW physics and for the LHC. 

But, how will we be able to establish that a new discovery made at the LHC is indeed the DM which fills the universe? First of all, dark matter, being electromagnetic and color neutral, would lead to events with missing-energy, when directly produced at the LHC. This prediction is robust and rather model-independent. 
Therefore, the study of the inclusive process ``missing transverse energy plus anything" is particularly interesting and, if an excess in the corresponding distributions is observed at the LHC, DM should be considered as a prime suspect. 

A much more ambitious way to identify the DM nature of the new discovery would be to reconstruct the thermal relic abundance from collider data. This requires measuring with high precision the masses and couplings of all the new particles associated to the DM sector. In very special cases this is possible at the LHC~\cite{dmlhc}, but the ILC appears to be more suited to this task~\cite{dmilc}, at least when the new particles are sufficiently light. Not only would the reconstruction of the thermal abundance be a convincing proof of the discovery of DM, but it would also allow us to reliably describe the history of the universe up to a temperature $T_f \simeq m_{DM}/20$. This would significantly extend the range of our knowledge of the early universe, which today can be viewed as firm only up to the nucleosynthesis temperature $T\simeq$~MeV. Of course the collider reconstruction of $\Omega_{DM}$ cannot be attempted  for DM candidates that have a non-thermal origin.

A third method is to combine collider data with results from non-accelerator DM searches. Indeed the information from different experimental strategies is totally complementary. While the LHC can detect new physics and give us information on masses and interactions, and the ILC can perform precision measurements on the relevant parameters, non-accelerator searches can link the discovered particle to its presence in the galactic halo. Direct detection methods can determine well the mass of the DM particle, but the total rate cannot be directly linked to the cosmological properties, because it depends on the details of the local halo density. Combining these results with indirect searches, one can derive information on the halo profile and the DM distribution.   

Finally, the LHC could identify special model-dependent features that would indirectly indicate the DM nature of the discovery. Take the example of the neutralino ($\chi$) in supersymmetry. While $\chi$ used to be a fully natural thermal DM candidate, the stringent limits from LEP and the precise determination of $\Omega_{DM}$ from WMAP~\cite{wmap} have made a quantitative difference. In the vast majority of the allowed parameter space, the relic abundance is either too large or too small. The correct value of $\Omega_{DM}$ can be obtained only under special conditions: a heavy spectrum with 1~TeV Higgsino or a 2.5~TeV Wino; nearly mass degenerate Bino and stau~\cite{bst} (or a light stop?) to allow coannihilation; nearly mass degenerate Bino--Higgsino or Bino--Wino states to obtain the ``well-tempered" condition~\cite{welltemp}; a new heavy Higgs with mass equal to $2m_\chi$ to allow for an $s$-channel resonance~\cite{heavyh}; slepton masses very close to their LEP bound to ensure a rapid Bino annihilation; non-thermal scenarios~\cite{nonth} with a reheat temperature after inflation close to the $\chi$ freeze-out temperature or with new particles with late decays into neutralinos. In all these possibilities the $\chi$ relic abundance has a very critical dependence on the underlying parameters, a property which may be viewed as theoretically unattractive, since supersymmetry cannot {\it naturally} predict the correct value of $\Omega_{DM}$. On the other hand, this situation can be viewed as experimentally more favorable, because identifying one of the previous conditions at the LHC will be a strong indication in favor of a DM discovery.

\section{Are there totally unexpected phenomena?}

I find that it would be, rather than embarrassing, very appealing for theorists if the LHC discovered some unexpected and unpredicted phenomenon. Experiments would lead the way to new ideas and new challenges. Nevertheless, theorists have done their best to avoid this outcome by proposing all kinds of new theories, and by making all kinds of possible (and sometimes impossible) predictions,
with good (or sometimes poor) motivations. Among the exotic proposals, let me mention three ideas that have attracted some attention.

In 1970, Lee and Wick~\cite{leewick} suggested that QED could be made finite by regulating the theory \`a la Pauli-Villars and by interpreting the associated ghosts as physical fields. They gave a prescription to eliminate any exponentially-growing mode in the final state (which actually corresponds to boundary conditions at future time and therefore explicit violation of causality), and claimed that no inconsistency arises in perturbation theory. However, no non-perturbative description using a path-integral formulation was found~\cite{gross}. It was recently suggested~\cite{newlee} that the hierarchy problem could be solved by canceling quadratic divergences with Pauli-Villars ghosts, giving them a Lee-Wick interpretation. I believe that it is still an interesting open question to demonstrate if these theories are really consistent. 

The concept of unparticle~\cite{unp} was introduced to describe hypothetical matter from a new scale-invariant sector of the theory. This proposal does not have any motivated links with physics at the Fermi scale, but the idea of having a conformal field theory (CFT) sector separated from the usual SM sector is present in other constructions. In this scheme, the two sectors communicate only through higher-dimensional operators suppressed by a large mass $M$. Below a scale $\Lambda$ ($< M$), the new sector becomes conformal and the original operators match onto new CFT operators whose scaling dimension determines the power of $\Lambda$ in the corresponding coefficient. No particle interpretation can be given to the states in the CFT. Nevertheless, one can compute the production of unparticles (described by an operator with a given Lorentz structure and scaling dimension $d_U$) from the decay of ordinary particles.  The decay width turns out to be equal to the rate of decay into a non-integer number $d_U$ of invisible particles. Moreover, the virtual exchange of unparticles with momentum $q$ gives an amplitude with imaginary part in the time-like region, for any $q^2>0$~\cite{unp2,changh}. These properties are not surprising in the context of a higher-dimensional theory, as it can be expected from the AdS/CFT correspondence. As previously discussed, the operator  in eq.~(\ref{effint}) is potentially most sensitive to the presence of hidden sectors. In this context, it gives a relevant coupling in the CFT and it directly affects the unparticle sector~\cite{newop}. Once the Higgs gets a vev, it provides a scale which breaks the conformal symmetry and the operator in eq.~(\ref{effint}) leads to an unparticle tadpole which can destabilize the theory. Higgs physics will be affected too.

Finally, 
various arguments have been given in ref.~\cite{dvali} to infer that, in a theory with  $N$ particles of mass $\Lambda$, the Planck mass must satisfy the bound $M_{Pl}^2 > N \Lambda^2$.  Making the bold suggestion that there are $10^{32}$ replica of the SM at the TeV scale, one can justify the big separation between the Fermi and Planck scale. The proposal may appear less shocking if one interprets the Large Extra Dimension solution to the hierarchy problem~\cite{add} also as a large-$N$ effect, due to the huge number of Kaluza-Klein graviton excitations.

\section{Conclusions: What is the mechanism of electroweak breaking?}

This question summarizes the main physics goal of the LHC. The investigation of the mechanism of EW breaking is a task certainly within the LHC energy reach. If the Higgs is discovered, the phenomenon giving rise to the Fermi scale will be unveiled. Measuring the properties of the Higgs will be fundamental to identify the new force responsible for EW-symmetry breaking and will give us insight on the underlying structure of the theory. Alternatives to the Higgs mechanism are theoretically interesting and phenomenologically possible. In the process of exploring the physics at the Fermi scale, new phenomena may emerge which can revolutionize our understanding of the space-time structure or of the nature of the fundamental interactions. If new physics associated with the hierarchy problem is discovered, it will be a new success of the concepts of reductionism and of symmetry. If the dark matter is identified, it will be a new triumph of the particle/cosmo connection.

\end{document}